\documentclass[10pt,conference]{IEEEtran}
\IEEEoverridecommandlockouts
\usepackage{cite}
\usepackage{subfig}
\usepackage{amsmath,amssymb,amsfonts}
\usepackage{algorithmic}
\usepackage{nopageno}
\usepackage{graphicx}
\usepackage{textcomp}
\usepackage{multirow}
\usepackage{xcolor}
\usepackage{array}
\usepackage{color,soul}
\usepackage[pass]{geometry}
\usepackage{hyperref}
\usepackage[acronyms,nonumberlist,nopostdot,nomain,nogroupskip]{glossaries}
\usepackage{tikz}
\usepackage[font=footnotesize]{caption}

\usetikzlibrary{positioning,shapes,shadows,arrows}
\definecolor{mycolor}{rgb}{1,0.2,0.2}
\definecolor{mycolor2}{rgb}{0.3,0.7,0.1}
\definecolor{mycolor10}{rgb}{0.,0.2,0.9}
\definecolor{mycolor3}{rgb}{0.9,0.9,0.9}
\tikzstyle{abstract}=[rectangle, draw=black, rounded corners, fill=mycolor, drop shadow,
        text centered, anchor=north, text=white, text width=2.5cm]
\tikzstyle{comment}=[rectangle, draw=black, rounded corners, fill=mycolor3, drop shadow,
        text centered, anchor=north, text=black, text width=2.5cm]
\tikzstyle{Sio}=[rectangle, draw=black, rounded corners, fill=mycolor2, drop shadow,
        text centered, anchor=north, text=white, text width=2.5cm]
\tikzstyle{Sio2}=[rectangle, draw=black, rounded corners, fill=mycolor10, drop shadow,
        text centered, anchor=north, text=white, text width=2.5cm]
\tikzstyle{myarrow}=[->, >=open triangle 90, thick]
\tikzstyle{line}=[-, thick]

\newacronym{3gpp}{3GPP}{3rd Generation Partnership Project}
\newacronym{adc}{ADC}{Analog to Digital Converter}
\newacronym{5g}{5G}{5th generation}
\newacronym{4g}{4G}{4th generation}
\newacronym{aimd}{AIMD}{Additive Increase Multiplicative Decrease}
\newacronym{am}{AM}{Acknowledged Mode}
\newacronym{amc}{AMC}{Adaptive Modulation and Coding}
\newacronym{aqm}{AQM}{Active Queue Management}
\newacronym{awgn}{AGWN}{Additive White Gaussian Noise}
\newacronym{balia}{BALIA}{Balanced Link Adaptation}
\newacronym{bdp}{BDP}{Bandwidth-Delay Product}
\newacronym{bf}{BF}{Beamforming}
\newacronym{cc}{CC}{Congestion Control}
\newacronym{cu}{CU}{Central Unit}
\newacronym{ecdf}{ECDF}{Empirical Cumulative Distribution Function}
\newacronym{cn}{CN}{Core Network}
\newacronym{cqi}{CQI}{Channel Quality Information}
\newacronym{cp}{CP}{Control Plane}
\newacronym{csirs}{CSI-RS}{Channel State Information - Reference Signal}
\newacronym{dc}{DC}{Dual Connectivity}
\newacronym{dce}{DCE}{Direct Code Execution}
\newacronym{dci}{DCI}{Downlink Control Information}
\newacronym{dl}{DL}{Downlink}
\newacronym{du}{DU}{Distributed Unit}
\newacronym{dmr}{DMR}{Deadline Miss Ratio}
\newacronym{dmrs}{DMRS}{DeModulation Reference Signal}
\newacronym{e2e}{E2E}{end-to-end}
\newacronym{ecn}{ECN}{Explicit Congestion Notification}
\newacronym{edf}{EDF}{Earliest Deadline First}
\newacronym{enb}{eNB}{evolved Node Base}
\newacronym{epc}{EPC}{Evolved Packet Core}
\newacronym{es}{ES}{Edge Server}
\newacronym{fdma}{FDMA}{Frequency Division Multiple Access}
\newacronym{fdd}{FDD}{Frequency Division Duplexing}
\newacronym[firstplural=Radio Access Technologies (RATs)]{rat}{RAT}{Radio Access Technology}
\newacronym{fs}{FS}{Fast Switching}
\newacronym{ftp}{FTP}{File Transfer Protocol}
\newacronym{gnb}{gNB}{Next Generation Node Base}
\newacronym{harq}{HARQ}{Hybrid Automatic Repeat reQuest}
\newacronym{hetnet}{HetNet}{Heterogeneous Network}
\newacronym{hh}{HH}{Hard Handover}
\newacronym{hol}{HOL}{Head-of-Line}
\newacronym{ia}{IA}{Initial Access}
\newacronym{ieee}{IEEE}{Institute of Electrical and Electronics Engineers}
\newacronym{ilp}{ILP}{Integer Linear Program}
\newacronym{imt}{IMT}{International Mobile Telecommunication}
\newacronym{iot}{IoT}{Internet of Things}
\newacronym{ldpc}{LDPC}{Low-Density Parity Check}
\newacronym{los}{LOS}{Line-of-Sight}
\newacronym{lte}{LTE}{Long Term Evolution}
\newacronym{m2m}{M2M}{Machine to Machine}
\newacronym{mac}{MAC}{Medium Access Control}
\newacronym{mc}{MC}{Multi-Connectivity}
\newacronym{mcs}{MCS}{Modulation and Coding Scheme}
\newacronym{mec}{MEC}{Mobile Edge Cloud}
\newacronym{mi}{MI}{Mutual Information}
\newacronym{mimo}{MIMO}{Multiple Input, Multiple Output}
\newacronym{mmwave}{mmWave}{millimeter wave}
\newacronym{mptcp}{MPTCP}{Multipath TCP}
\newacronym{mr}{MR}{Maximum Rate}
\newacronym{mss}{MSS}{Maximum Segment Size}
\newacronym{mtd}{MTD}{Machine-Type Device}
\newacronym{mtu}{MTU}{Maximum Transmission Unit}
\newacronym{nfv}{NFV}{Network Function Virtualization}
\newacronym{nlos}{NLOS}{Non-Line-of-Sight}
\newacronym{nlosv}{NLOSv}{Vehicle Non-Line-of-Sight}
\newacronym{nr}{NR}{New Radio}
\newacronym{ofdm}{OFDM}{Orthogonal Frequency Division Multiplexing}
\newacronym{pdcch}{PDCCH}{Physical Downlonk Control Channel}
\newacronym{pdcp}{PDCP}{Packet Data Convergence Protocol}
\newacronym{pdsch}{PDSCH}{Physical Downlink Shared Channel}
\newacronym{pdu}{PDU}{Packet Data Unit}
\newacronym{pf}{PF}{Proportional Fair}
\newacronym{pgw}{PGW}{Packet Gateway}
\newacronym{phy}{PHY}{Physical}
\newacronym{pbch}{PBCH}{Physical Broadcast Channel}
\newacronym[plural=\gls{mme}s,firstplural=Mobility Management Entities (MMEs)]{mme}{MME}{Mobility Management Entity}
\newacronym{prb}{PRB}{Physical Resource Block}
\newacronym{pss}{PSS}{Primary Synchronization Signal}
\newacronym{pscch}{PSCCH}{Physical Sidelink Control Channel}
\newacronym{pucch}{PUCCH}{Physical Uplink Control Channel}
\newacronym{pusch}{PUSCH}{Physical Uplink Shared Channel}
\newacronym{rach}{RACH}{Random Access Channel}
\newacronym{ran}{RAN}{Radio Access Network}
\newacronym{red}{RED}{Random Early Detection}
\newacronym{rf}{RF}{Radio Frequency}
\newacronym{rlc}{RLC}{Radio Link Control}
\newacronym{rlf}{RLF}{Radio Link Failure}
\newacronym{rrc}{RRC}{Radio Resource Control}
\newacronym{rrm}{RRM}{Radio Resource Management}
\newacronym{rr}{RR}{Round Robin}
\newacronym{rs}{RS}{Remote Server}
\newacronym{rsrp}{RSRP}{Reference Signal Received Power}
\newacronym{rss}{RSS}{Received Signal Strength}
\newacronym{rtt}{RTT}{Round Trip Time}
\newacronym{rw}{RW}{Receive Window}
\newacronym{rx}{RX}{Receiver}
\newacronym{sa}{SA}{standalone}
\newacronym{sack}{SACK}{Selective Acknowledgment}
\newacronym{sap}{SAP}{Service Access Point}
\newacronym{sc}{SC}{Single Carrier}
\newacronym{sch}{SCH}{Secondary Cell Handover}
\newacronym{scoot}{SCOOT}{Split Cycle Offset Optimization Technique}
\newacronym{sdma}{SDMA}{Spatial Division Multiple Access}
\newacronym{sinr}{SINR}{Signal to Interference plus Noise Ratio}
\newacronym{sl}{SL}{Sidelink}
\newacronym{sm}{SM}{Saturation Mode}
\newacronym{snr}{SNR}{Signal-to-Noise-Ratio}
\newacronym{son}{SON}{Self-Organizing Network}
\newacronym{ss}{SS}{Synchronization Signal}
\newacronym{srs}{SRS}{Sounding Reference Signal}
\newacronym{sss}{SSS}{Secondary Synchronization Signal}
\newacronym{tb}{TB}{Transport Block}
\newacronym{tcp}{TCP}{Transmission Control Protocol}
\newacronym{tdd}{TDD}{Time Division Duplexing}
\newacronym{tdma}{TDMA}{Time Division Multiple Access}
\newacronym{tfl}{TfL}{Transport for London}
\newacronym{tm}{TM}{Transparent Mode}
\newacronym{trp}{TRP}{Transmitter Receiver Pair}
\newacronym{tti}{TTI}{Transmission Time Interval}
\newacronym{ttt}{TTT}{Time-to-Trigger}
\newacronym{tx}{TX}{Transmitter}
\newacronym{ue}{UE}{User Equipment}
\newacronym{ul}{UL}{Uplink}
\newacronym{uml}{UML}{Unified Modeling Language}
\newacronym{um}{UM}{Unacknowledged Mode}
\newacronym{utc}{UTC}{Urban Traffic Control}
\newacronym{vm}{VM}{Virtual Machine}
\newacronym{rsrq}{RSRQ}{Reference Signal Received Quality}
\newacronym{rssi}{RSSI}{Received Signal Strength Indicator}
\newacronym{crs}{CRS}{Cell Reference Signal}
\newacronym{nsa}{NSA}{Non Stand Alone}
\newacronym{mrdc}{MR-DC}{Multi \gls{rat} \gls{dc}}
\newacronym{endc}{EN-DC}{E-UTRAN-\gls{nr} \gls{dc}}
\newacronym{5gc}{5GC}{5G Core}
\newacronym{si}{SI}{Study Item}
\newacronym{iab}{IAB}{Integrated Access and Backhaul}
\newacronym{sbs}{SeBaSi}{Self-Backhauling-Simulator}
\newacronym{wf}{WF}{Wired-first}
\newacronym{hqf}{HQF}{Highest-quality-first}
\newacronym{pa}{PA}{Position-aware}
\newacronym{mlr}{MLR}{Maximum-local-rate}
\newacronym{wbf}{WBF}{Wired Bias Function}
\newacronym{mib}{MIB}{Master Information Block}
\newacronym{sib}{SIB}{Secondary Information Block}
\newacronym{rnti}{RNTI}{Radio Network Temporary Identifier}
\newacronym{dft}{DFT}{Discrete Fourier Transform}
\newacronym{kpi}{KPI}{Key Performance Indicator}
\newacronym{ppp}{PPP}{Poisson Point Process}
\newacronym{v2v}{V2V}{Vehicle-to-Vehicle}
\newacronym{wave}{WAVE}{Wireless Access in Vehicular Environments}
\newacronym{udp}{UDP}{User Datagram Protocol}
\newacronym{upa}{UPA}{Uniform Planar Array}
\newacronym{fec}{FEC}{Forward Error Correction}
\newacronym{v2x}{V2X}{Vehicle-To-Everything}
\newacronym{psfch}{PSFCH}{Physical Sidelink Feedback Channel}
\newacronym{pssch}{PSSCH}{Physical Sidelink Shared Channel}
\newacronym{csma}{CSMA}{Carrier Sense Multiple Access}
\newacronym{v2n}{V2N}{Vehicle-to-Network}
\newacronym{wlan}{WLAN}{Wireless Local Area Network}
\newacronym{cav}{CAV}{Connected and Autonomous Vehicle}
\newacronym{v2i}{V2I}{Vehicle-to-Infrastructure}
\newacronym{d2d}{D2D}{Device-to-Device}
\newacronym{c-its}{C-ITS}{Connected Intelligent Transportation System}
\newacronym{fr2}{FR2}{Frequency Range 2}
\newacronym{bs}{BS}{Base Station}
\newacronym{sdu}{SDU}{Service Data Unit}
\newacronym{csi}{CSI}{Channel State Information}
\newacronym{scs}{SCS}{Subcarrier Spacing}
\newacronym{sumo}{SUMO}{Simulation of Urban MObility}
\newacronym{prr}{PRR}{Packet Reception Ratio}
\newacronym{edca}{EDCA}{Enhanced Distribution Channel Access}
\newacronym{sdap}{SDAP}{Service Data Adaptation Protocol}
\newacronym{thz}{THz}{terahertz}
\newacronym{scm}{SCM}{Spatial Channel Model}
\newacronym{vr}{VR}{Virtual Reality}
\newacronym{6g}{6G}{6th generation}
\newacronym{qos}{QoS}{Quality of Service}
\newacronym{uav}{UAV}{unmanned Aerial Vehicles}
\newacronym{bap}{BAP}{Backhaul Adaptation Protocol}
\newacronym{ns3}{ns-3}{Network Simulator 3}
\newacronym{rl}{RL}{Reinforcement Learning}
\newacronym{ris}{RIS}{Reconfigurable Intelligent Surface}

\def\BibTeX{{\rm B\kern-.05em{\sc i\kern-.025em b}\kern-.08em
    T\kern-.1667em\lower.7ex\hbox{E}\kern-.125emX}}
\begin{document}
\newcommand{\coolname}{\gls{sbs}}
\title{Demo:[\coolname] system-level Integrated Access and Backhaul simulator for self-backhauling 
}
\author{
\IEEEauthorblockN{Amir Ashtari Gargari\IEEEauthorrefmark{1}, Matteo Pagin\IEEEauthorrefmark{1}, Andrea Ortiz\IEEEauthorrefmark{2}, Nairy Moghadas Gholian\IEEEauthorrefmark{4}, Michele Polese\IEEEauthorrefmark{7}, Michele Zorzi\IEEEauthorrefmark{1}}
\IEEEauthorblockA{\IEEEauthorrefmark{1}Department of Information Engineering, University of Padova, Italy, \{amirashtari, paginmatte, zorzi\}@dei.unipd.it}
\IEEEauthorblockA{\IEEEauthorrefmark{2}Communications Engineering Lab, TU Darmstadt, Germany, a.ortiz@nt.tu-darmstadt.de}
\IEEEauthorblockA{\IEEEauthorrefmark{4}Wireless Communications and Sensing Lab (WISE), TU Darmstadt, Germany, ngholian@wise.tu-darmstadt.de}
\IEEEauthorblockA{\IEEEauthorrefmark{7}Institute for the Wireless Internet of Things, Northeastern University, Boston, MA, m.polese@northeastern.edu}
}


\maketitle

\begin{abstract}
\gls{mmwave} and sub-\gls{thz} communications have the potential of increasing mobile network throughput drastically. However, the challenging propagation conditions experienced at \gls{mmwave} and beyond frequencies can potentially limit the range of the wireless link down to a few meters, compared to up to kilometers for sub-6GHz links. Thus, increasing the density of base station deployments is required to achieve sufficient coverage in the \gls{ran}.
To such end, \gls{3gpp} introduced wireless backhauled base stations with \gls{iab}, a key technology to achieve dense 
networks while preventing the need for costly fiber deployments. 

In this paper, we introduce \coolname, a system-level simulator for \gls{iab} networks, and demonstrate its functionality by simulating \gls{iab} deployments in Manhattan, New York City and Padova. Finally, we show how \coolname~can represent a useful tool for the performance evaluation of self-backhauled cellular networks, thanks to its high level of network abstraction, coupled with its open and customizable design, which allows users to extend it to support novel technologies such as \glspl{ris}.
\end{abstract}

\begin{IEEEkeywords}
\gls{mmwave}, \gls{iab}, self-backhauling, wireless backhaul, sionna
\end{IEEEkeywords}

\section{Introduction}

\glsunset{nr}

5G mobile networks introduced the support for \gls{mmwave} communications, with a further expansion towards sub-\gls{thz} envisioned for \gls{6g} networks. This progressive shift from sub-6~GHz frequencies towards the upper portion of the spectrum represents the main technology enabler towards achieving multi-Gbps mobile throughput. Nevertheless, \gls{mmwave} and sub-\gls{thz} frequencies are affected by high propagation and penetration losses, as well as by a marked susceptibility to blockage, which degrade the reliability and capacity of wireless networks operating in this portion of the spectrum~\cite{9269934,gargari2021full}. 
To mitigate these unfavorable propagation characteristics, it is paramount to maximize the \gls{los} coverage, and to increase the density of base station deployments with respect to traditional sub-6~GHz cellular networks. 
In turn, to make ultra-dense deployments a viable option from both a logistic and an economic standpoint, the \gls{3gpp} has standardized an extension of \gls{5g} \gls{nr} known as \gls{iab}\cite{3gpp.38.300, 3gpp_38_174}. The latter leverages the \gls{du}/\gls{cu} split to introduce \glspl{gnb} with wireless backhauling capabilities, i.e., \gls{iab} nodes, thus reducing the need for fiber drops. The \gls{iab} nodes eventually terminate the various 5G NR interfaces at a \gls{gnb} connected by fiber to the \gls{cn} and the Internet, i.e., the \gls{iab} donor.

In this context, 
the research community has been studying how to optimize radio resource allocation, scheduling, route selection, topology construction, and deployment planning\cite{kwon2019joint,pizzo2017optimal}. Given the lack of access to actual \gls{5g} (and beyond) network deployments, previous research efforts relied heavily on homegrown physical layer simulators to evaluation performance\cite{8876705}. 
However, these simulators usually feature a heavily simplified model of actual \gls{iab} deployments, since they introduce strong assumptions in the upper layers of the protocol stack. Therefore, they are incapable of capturing the real network dynamics. Similarly, existing system-level simulators are outdated, and thus also fail to properly model a Rel. 17 \gls{iab} network\cite{polese2018end}.
Moreover, an experimental evaluation is prohibitive since researchers usually do not have access to commercial deployments at scale. 
To fill this gap, we introduce \coolname, an \gls{iab} simulator 
which accurately models large-scale wireless backhauled deployments.
In this demo, we describe \coolname, and showcase examples of different \gls{iab} cellular networks, for which we report system-level \glspl{kpi}. 

This paper is organized as follows. In Sec.~\ref{sec:System}, we describe the system model. Then, we introduce the proposed system-level simulator in Sec.~\ref{sec:coolname}, and describe the contents of the demo in Sec.~\ref{sec:Demo}. Finally, we discuss possible future extension of our simulator in Sec.~\ref{SEC:future}. 

\section{System Model}
\label{sec:System}
In this work, we consider a \gls{tdma} cellular system where both self-backhauled and wired base stations exchange data with the \glspl{ue}. In accordance with \gls{3gpp} terminology, we refer to the former base stations as IAB-nodes (BS-nodes) and to the latter as IAB-donors (BS-donors). The IAB-nodes exchange data with the \gls{cn} and the Internet via a wireless multi-hop connection to an IAB-donor. 

We assume that the IAB-nodes incorporate two \gls{rf} chains. One \gls{rf} chain is reserved for communication with cellular users (access network), while the other is utilized for self-backhauling. 
In line with the \gls{3gpp} standard~\cite{3gpp_38_174}, we assume half-duplex and in band self-backhauling, such that in each time slot the IAB-nodes can either transmit, receive, or remain idle. 
Without loss of generality, in this analysis, we focus on uplink traffic only. 

\section{\coolname}
\label{sec:coolname}
\coolname~is a Python system-level simulator, built on top of the open-source \gls{5g} and \gls{6g} physical layer simulator Sionna\textsuperscript{TM}\cite{hoydis2022sionna}, which models \gls{3gpp} Rel. 17 \gls{iab} cellular networks. 
To introduce self-backhauling functionalities in Sionna, we have implemented a number of system-level features. These extensions, which we describe more in detail in~\cite{INFOCOMSIM}, 
comprise a \gls{mac}-level scheduler, layer-2 buffers and backhaul path selection algorithms. 
Moreover, to better align Sionna's physical layer to that of \gls{5g}-\gls{nr}, we also implemented \gls{5g}-\gls{nr} procedures such as codebook-based beamforming and \gls{sinr} computation. 
Additionally, in~\cite{IABwons} we further extend our simulator to support sub-\gls{thz} links in the backhaul, with the goal of 
providing a performance evaluation of the potential of sub-\gls{thz} frequencies for 6G IAB.

\begin{figure}\label{bigpicture}\centering
\resizebox{.47\textwidth}{!}{
\begin{tikzpicture}[node distance=2cm,thick,scale=1, every node/.style={scale=1}]
    \node (Item) [abstract, rectangle split parts=2]
        {
            \textbf{Simulator}
        };
    \node (AuxNode01) [text width=4cm, below=of Item, xshift=1cm] {};
    \node (Component) [abstract, rectangle split, rectangle split parts=2,  below=of Item, xshift=1.5cm, yshift=0.7cm]
        {
        
            \textbf{nonPHY}
            \nodepart{second} 
        };
    \node (class) [abstract, rectangle split, rectangle split parts=2, left=of Component, xshift=-0.5cm]
        {
            \textbf{DataTypes}
            \nodepart{second} sub Class()
        };
    \node (System) [abstract, rectangle split, rectangle split parts=2, right=of Component ,xshift=2cm]
        {
            \textbf{PHY}
            \nodepart{second} Physical
        };
    \node (AuxNode02) [text width=0.5cm, below=of Component] {};
    \node (AuxNode10) [text width=0.5cm, below=of class] {};
     \node (iab) [abstract, rectangle split, rectangle split parts=2, below=of class, xshift=0cm, yshift=1.4cm]
        {
            \textbf{iab()}
            \nodepart{second}DataType
        }; 

    \node (iabClasses) [comment, rectangle split, rectangle split parts=2, below=0.2cm of iab, text justified]
        {
            \textbf{Classes}
            \nodepart{second}iab\_id \newline cell\_id \newline location \newline Antenna \& BF  \newline Tx\_Buffer \newline Rx\_Buffer 
        }; 
     \node (network) [abstract, rectangle split, rectangle split parts=2, left=of class, xshift=1.5cm]
        {
            \textbf{Setup()}
            \nodepart{second}Class
        }; 

    \node (netClasses) [comment, rectangle split, rectangle split parts=2, below=0.2cm of network, text justified]
        {
            \textbf{Classes}
            \nodepart{second} Topology() \newline simulation\_mode \newline run\_time \newline Access() \newline Backhaul()
        }; 
    \node (packet) [abstract, rectangle split, rectangle split parts=2, below=of netClasses, yshift=1.3cm]
        {
            \textbf{packet()}
            \nodepart{second}DataType
        }; 

    \node (packetClasses) [comment, rectangle split, rectangle split parts=2, below=0.2cm of packet, text justified]
        {
            \textbf{Classes}
            \nodepart{second}packet\_id \newline packet\_size \newline frame\_id \newline UE\_id \newline Cell\_id \newline Delay \newline passed\_path\_list 
        };

    \node (path) [abstract, rectangle split, rectangle split parts=2, below=of iabClasses, yshift=1.7cm]
        {
            \textbf{path()}
            \nodepart{second}DataType
        }; 

    \node (pathInstant) [comment, rectangle split, rectangle split parts=2, below=0.2cm of path, text justified]
        {
            \textbf{Classes}
            \nodepart{second}path\_id \newline from\_IAB \newline to\_IAB \newline SINR \newline path\_rate \newline path\_equipped 
        };

    \node (Sensor) [abstract, rectangle split, rectangle split parts=2, left=of AuxNode02, xshift=1.8cm]
        {
            \textbf{DataFlow}
            \nodepart{second} 
        };

    \node (Part) [abstract, rectangle split, rectangle split parts=2, right=of AuxNode02, xshift=-1.8cm]
        {
            \textbf{Control Plane}
            \nodepart{second}
        };
        

    \node (Temperature) [abstract, rectangle split, rectangle split parts=2, below=of Sensor, xshift=0cm, yshift=1cm]
        {
            \textbf{Buffer}
            \nodepart{second}RLC-like
        };
    \node (TemperatureClasses) [comment, rectangle split, rectangle split parts=2, below=0.2cm of Temperature, text justified]
        {
            \textbf{Classes}
            \nodepart{second}RX DU RLC \newline TX MT RLC
        };

    \node (Pump) [abstract, rectangle split, rectangle split parts=2, below=of Part, xshift=0cm,yshift=1cm]
        {
            \textbf{BAP-like}
            \nodepart{second} 
        };
    \node (PumpClasses) [comment, rectangle split, rectangle split parts=2, below=0.2cm of Pump, text justified]
        {
            \textbf{Classes}
            \nodepart{second}3GPP Standard TS 38.340
        };
     \node (pathsel) [abstract, rectangle split, rectangle split parts=2, below=of PumpClasses, xshift=0cm,yshift=1.7cm]
        {
            \textbf{Path Selection Policy}
            \nodepart{second}
        };    

    \node (pathClasses) [comment, rectangle split, rectangle split parts=2, below=0.2cm of pathsel, text justified]
        {
            \textbf{Classes}
            \nodepart{second} \coolname \newline SCAROS \newline MLR
        };    
    \node (Valve) [abstract, rectangle split, rectangle split parts=2, right=of pathClasses, xshift=0cm, yshift=1.85cm]
        {
            \textbf{Scheduler}
            \nodepart{second} 
        };

    \node (ValveClasses) [comment, rectangle split, rectangle split parts=2, below=0.2cm of Valve, text justified]
        {
            \textbf{Classes}
            \nodepart{second} RoundRobin()
        };

    \node (AuxNode05) [below=of System] {};
    \node (CoolingSystem) [Sio, rectangle split, rectangle split parts=2, left=of AuxNode05, xshift=2cm]
        {
            \textbf{Sionna}
            \nodepart{second} \cite{hoydis2022sionna}
        };
    \node (CoolingLoop) [abstract, rectangle split, rectangle split parts=2, right=of AuxNode05, xshift=-2cm]
        {
            \textbf{Extended Modules}
            \nodepart{second} 
        };
    \node (CoolingLoop2) [abstract, rectangle split, rectangle split parts=2, right=of CoolingLoop, xshift=-1cm]
        {
            \textbf{Sub-THz Channel Traces}
            \nodepart{second} \cite{IABwons}
        };
    \node (CoolingSystemClasses) [comment, rectangle split, rectangle split parts=2, below=0.2cm of CoolingSystem, text justified]
        {
            \textbf{Classes}
            \nodepart{second}
            \newline 3GPP TR 38.901 Channel models
            \newline MU-MIMO
            \newline OFDM
            \newline 5G-NR

        };
    \node (CoolingSystemClasses2) [comment, rectangle split, rectangle split parts=2, below=0.2cm of CoolingLoop2, text justified]
        {
            \textbf{Classes}
            \nodepart{second}
            \newline Physical Based THz Channel models
            \newline 140GHz
            \newline horn Antenna

        };
    \node (CoolingLoopClasses) [comment, rectangle split, rectangle split parts=2, below=0.2cm of CoolingLoop, text justified]
        {
            \textbf{Classes}
            \nodepart{second}Beamforming\newline ICI \& SINR
        };
    
    \draw[myarrow] (Component.north) -- ++(0,0.8) -| (Item.south);
    \draw[line] (Component.north) -- ++(0,0.8) -| (System.north);
    \draw[line] (network.north) -- ++(0,0.8) -| (Item.south);
    \draw[myarrow] (class.north) -- ++(0,0.8) -| (Item.south);

    \draw[myarrow] (Sensor.north) -- ++(0,0.8) -| (Component.south);
    \draw[line] (Sensor.north) -- ++(0,0.8) -| (Part.north);

    \draw[line] (Pump.east) -- ++(0.2,0);
     ([xshift=-1cm]Sensor.south);
     \draw[myarrow] (Temperature.north) -- ++(0,0.8) -| (Sensor.south);

    \draw[line] (iab.west) -- ++(-0.2,0);
    \draw[line] (packet.east) -- ++(0.3,0);
    \draw[myarrow] (path.west) -- ++(-0.2,0) -- ([yshift=-0.5cm, xshift=-0.2cm] class.south west) -|([xshift=0cm]class.south);
    

     ([xshift=-1cm]Part.south);
    \draw[myarrow] (pathsel.east) -- ++(0.2,0) -- ([yshift=0.5cm, xshift=0.2cm] Pump.north east) -|
     ([xshift=0cm]Part.south);
     
     \draw[line] (pathsel.east) -- ++(0,0.) -| (Valve.west);
    \draw[myarrow] (CoolingSystem.north) -- ++(0,0.8) -| (System.south);
    \draw[line] (CoolingSystem.north) -- ++(0,0.8) -| (CoolingLoop.north);
    \draw[line] (CoolingSystem.north) -- ++(0,0.8) -| (CoolingLoop2.north);

\end{tikzpicture}
}
\caption{Overall design of \coolname. The red blocks represent our additions to the baseline simulator, i.e., Sionna~\cite{hoydis2022sionna}.}
\label{fig:simulator_design}
\end{figure}
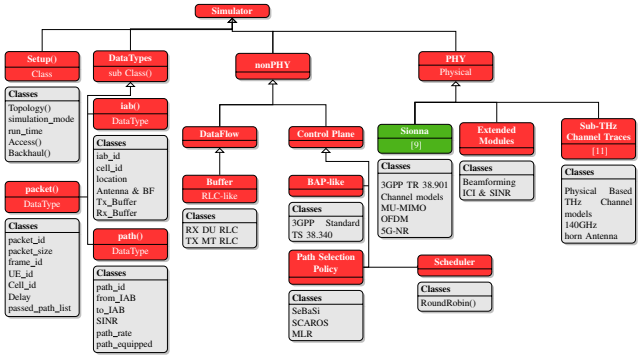

Fig.~\ref{fig:simulator_design} depicts the general structure of \coolname. For the \gls{mmwave} channel, we rely on the \gls{3gpp} TR.~38.901 channel model provided by Sionna. Additionally, we model sub-THz channels using \gls{ns3}~Terasim\cite{hossain2018terasim}, and produce traces which are imported into our simulator. For the upper layers, we introduce the \gls{bap} layer, to handle routing within the wireless backhaul network~\cite{3gpp_38_340}, a \gls{mac}-level scheduler that operates in a \gls{tdma} fashion, and hop-by-hop \gls{rlc} channels for modeling layer-2 buffering and data transmission.

\coolname, which we make publicly available\footnote{\url{https://github.com/TUDA-wise/safehaul\_infocom2023}}, 
allows users to configure most simulation parameters, such as the simulation's runtime and mode, the packet size, and the source rate (either per UE or system-wise). The considered simulation modes are \textit{run} mode and \textit{debug} mode, with the latter providing additional control signals and related information. 
Moreover, users can customize the scenario by choosing the number and location of \glspl{ue} and base stations, and the \gls{iab} topology, i.e., the wireless backhaul links among \glspl{gnb}.
For the backhaul scheduler, i.e., 
the entity which dictates which backhaul links to schedule during each time slot, users can either define custom policies, or choose among: SCAROS~\cite{8876705}, MLR~\cite{9500058}, Safehaul~\cite{INFOCOMSIM}, and SINR-based~\cite{IABwons}. In addition, the links can be configured to operate either at mmWave, sub-THz, or a combination of the two frequencies. 

The simulator outputs an extensive set of system-level \glspl{kpi}, such as end-to-end latency, throughput, and packet drop rate. Each of these metrics can be displayed per \gls{iab} node, or for the entire network. In addition, we also make available internal and/or lower layer metrics such as the generation and arrival time of each packet, destination UE, and its backhaul path. Furthermore, we report the load on each IAB node per each time step and for both the access and backhaul interfaces.

\section{Demo Description}
\label{sec:Demo}
In this demonstration, we simulate in \coolname~two cellular deployments whose topologies mimic those of the cities of New York and Padova, respectively. To such end, we gathered \gls{4g} \gls{enb} locations from the actual deployments of three mobile network operators, and considered them as either \gls{5g} \gls{iab} nodes or donors. 

\begin{figure*}
\centering
    \subfloat[Per-UE end-to-end latency]{
    \includegraphics[width=.3\linewidth,height=0.48\columnwidth]{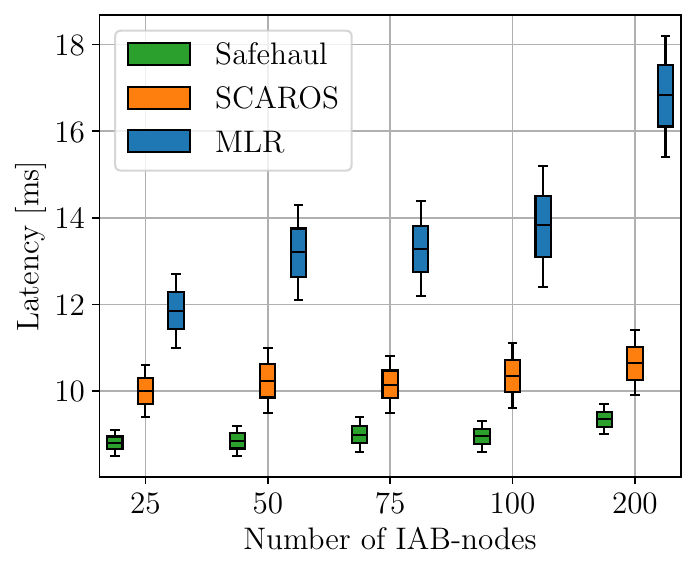}
    \label{fig:avgE2EDelay_s2}
    }
    \subfloat[Per-UE throughput]{
    \includegraphics[width=.3\linewidth,height=0.48\columnwidth]{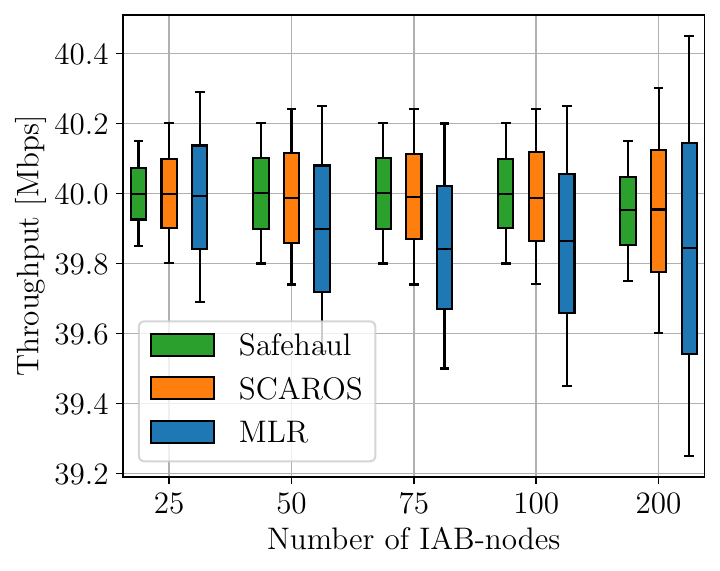}
    \label{fig:avgTput_s2}
    }
    \subfloat[Per-UE packet drop rate]{
    \includegraphics[width=.3\linewidth,height=0.48\columnwidth]{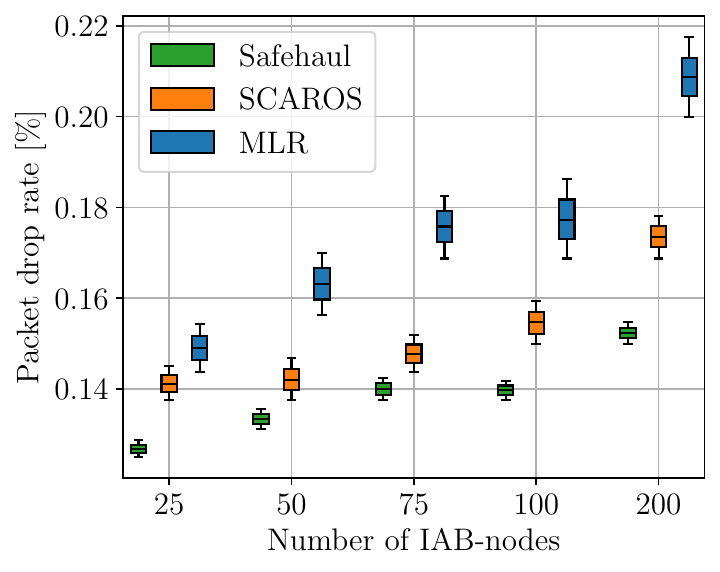}
    \label{fig:avgDropRate_s2}
    }
  
   \caption{Example of network \glspl{kpi} which can be obtained from \coolname, for $\{25, 50, 75, 100, 200\}$ \gls{iab}-nodes and $40$ Mbps per-UE source rate\cite{INFOCOMSIM}}
  \label{fig:avgNetPerfomance_s2}
  \vspace*{-3mm}
\end{figure*}

The overall objective of the demo is to provide examples of how to interact with \coolname, with different \gls{iab} 
deployments, and how to tune simulation parameters such as the system source rate and the frequency spectrum, by either defining new scenarios or by using the built-in examples of \coolname.
Moreover, we provide examples of \coolname's output traces by displaying in real-time \gls{kpi} metrics and network routing information, such as the one depicted in Fig.~\ref{fig:avgNetPerfomance_s2}.

Specifically, we first demonstrate how to work with \coolname~by instantiating the Manhattan, New York example. 
Then, we instantiate the Padova, Italy topology, and we demonstrate the performance of the backhaul scheduler Safehaul~\cite{INFOCOMSIM} in both scenarios. Safehaul is a risk-averse \gls{rl} solution to ensure reliability in IAB mmWave networks.

We conclude the demonstration by discussing future extensions and test-case scenarios. 

\section{Future work}
\label{SEC:future}

The deployment of \glspl{iab} is becoming increasingly important for seamless connectivity at \gls{mmwave} frequencies as the number of \glspl{ue} and their anticipated \gls{qos} grows. The operational cost for adding  more \glspl{iab} to meet the network's increased demand is high. Adding more of them to mitigate this issue is therefore not always possible.

A more cost-effective solution is to use \glspl{ris}. \glspl{ris} are energy-efficient smart surfaces that can change the direction of the impinging signal to the desired locations. The propagation characteristics may be enhanced in significant ways thanks to RISs\cite{trichopoulos2022design}. That's because \gls{mmwave} frequencies have high propagation along penetration loss, which increases with blockages, especially in urban scenarios. Using \glspl{ris} to their full extent can thus greatly avoid this problem. 

Yet, the coexistence of \glspl{iab} and \glspl{ris} presents particular challenges. The coordination between the \gls{ris} phase shifters along with the setting of \glspl{iab} would require solving complex joint optimization problems. Also, the placement of \glspl{ris} along \glspl{iab} in a network  would require more information on the \glspl{ue} location. Moreover, \glspl{ris} can interfere with other components of a cellular network, therefore, it's essential to carefully manage them such that the interference is avoided.
Because of the aforementioned factors, system-level simulation is required to verify \gls{ris}'s potential in actual situations. Both \gls{iab} and \gls{ris} serve as relays to increase the range of communication; \gls{iab} is active, whereas \gls{ris} is passive. Both of them function as relay elements from a software perspective, but channel modeling and beamforming in \gls{ris} are difficult\cite{noh2022channel}. We will attempt to integrate \gls{ris} into \coolname~using the new raytracing feature of Sionna. In summary, extending \coolname~to \gls{ris}-assisted scenarios will open a new direction in the future of \gls{mmwave} networks, however, several obstacles have to be addressed in order to fully exploit its potential.

\section{conclusion}

We have described and showcased \coolname~a system level \gls{iab} simulator and an example of self-backhauling in Manhattan, New York City.  In the demonstration, we have used Safehaul\cite{INFOCOMSIM} self-backhauling scheduler in \coolname~to evaluate the \gls{iab} network based on \gls{kpi}. Thanks to the open-source development of the tool, researchers can utilize new scenarios or extend the simulator to support novel technologies such as \gls{ris}.
\section*{Acknowledgment}

This paper is partially supported by EU H2020 MSCA ITN project MINTS (grant no. 861222), the Collaborative Research Center 1053 MAKI and the BMBF project Open6GHub (Nr. 16KISK014).

\bibliographystyle{IEEEtran}
\bibliography{IEEEabrv, bibl}
\end{document}